\documentclass[nohyper,notoc]{JHEP3}

\usepackage{amsmath,euscript,array,amssymb,cite,bm} 
\usepackage{graphicx} 
\usepackage{subfigure}
\usepackage{axodraw}                                      
\def\Tr{{\rm Tr}}
\def\sst{\scriptscriptstyle}
\def\det{{\rm det}}

\def\Dbarslash{\,\,{\raise.15ex\hbox{/}\mkern-12mu {\bar\D}}}
\def\Dslash{\,\,{\raise.15ex\hbox{/}\mkern-12mu \D}}
\def\delslash{\,\,{\raise.15ex\hbox{/}\mkern-9mu \partial}}
\def\delbarslash{\,\,{\raise.15ex\hbox{/}\mkern-9mu {\bar\partial}}}

\def\hf{{\textstyle{1\over2}}}

\def\D{{\cal D}}
\def\Dbarslash{\,\,{\raise.15ex\hbox{/}\mkern-12mu {\bar\D}}}
\def\delslash{\,\,{\raise.15ex\hbox{/}\mkern-9mu \partial}}
\def\Dslash{\,\,{\raise.15ex\hbox{/}\mkern-12mu \D}}

\def\={\, =\, }
\def\+{\, +\, }
\def\-{\, -\, }

\newcommand{\be}{\begin{equation}}
\newcommand{\ee}{\end{equation}}
\def\bea{\begin{eqnarray}}
\def\eea{\end{eqnarray}}

\def\Nf{{N_{f}}}
\def\Nc{{N_{c}}}
\def\p{\varphi}
\def\pt{\tilde\varphi}
\def\uno{\mbox{1 \kern-.59em {\rm l}}} 
\def\vplus{|{\rm vac}\rangle_+}
\def\vzero{|{\rm vac}\rangle_0}

\title{Why the early Universe preferred the non-supersymmetric vacuum: Part II}

\author{Steven A.~Abel, Joerg Jaeckel and Valentin V.~Khoze\\
Centre for Particle Theory and IPPP, University of Durham,
Durham, DH1 3LE, UK\\
{\tt s.a.abel@durham.ac.uk,}\,  
{\tt jjaeckel@mail.desy.de,}\, 
{\tt valya.khoze@durham.ac.uk}}

\abstract{It was recently shown in hep-th/0610334 that in the context of the ISS models
with a metastable supersymmetry breaking vacuum, thermal effects generically drive the Universe to the 
metastable vacuum even if it began after inflation  in the supersymmetry-preserving one. We continue 
this programme and specifically take into account two new effects. First is the effect of the mass-gap
of the gauge degrees of freedom in the confining supersymmetry preserving vacua, and second, is the 
effect of the back reaction of the MSSM sector on the SUSY breaking ISS sector. It is shown that, even 
though the mass-gap is parametrically 
smaller than the $\langle\p\rangle$, $\langle\pt\rangle$ vevs, it drastically reduces the 
temperature required for the Universe to be driven to the 
metastable vacuum: essentially any temperature larger 
than the supersymmetry breaking scale $\mu$ is sufficient. On the other hand we also find that
any reasonable transmission of SUSY breaking to the MSSM sector has no effect 
on the vacuum transitions to, and the stability of the SUSY breaking vacuum. 
We conclude that for these models the early Universe does end up in the SUSY breaking vacuum.}
 
\preprint{{\tt hep-th/0611130}
\\IPPP/06/78\\
DCPT/06/156
}

\begin{document}

\section{Introduction}

There is renewed interest in field-theoretical settings where a supersymmetric 
extension of the Standard Model (the MSSM for example) is coupled to another sector 
that contains supersymmetry preserving as well as a long-lived metastable supersymmetry 
breaking vacua (MSB). Intriligator, Seiberg and Shih (ISS) have argued that such 
metastable supersymmetry breaking can be a viable, calculable and remarkably simple alternative
to the usual scenarios of dynamical supersymmetry breaking \cite{ISS}. If the Universe started
in the metastable vacuum, then the rate of escape to the supersymmetry preserving
true vacuum is parametrically small in the ISS scenario and the lifetime of the metastable vacuum can easily be 
much longer than the age of the Universe. Supersymmetry remains broken without paying the 
usual high price for arranging the hidden sector to contain no supersymmetric ground states.
This opens a possibility for the supersymmetry of the full theory (including the MSSM sector) 
to be broken by the metastable vacuum of the MSB sector.

An important question that immediately follows is 
why did the Universe start from the non-supersymmetric vacuum in the first place \cite{ACJK}?
In our earlier paper \cite{ACJK} we provided a simple and generic explanation: thermal effects drive the Universe
to the supersymmetry breaking vacuum even if it ends up after inflation in the supersymmetry preserving ground state.
This happens for a large class of models, of which the ISS model is an example, that satisfy the following conditions:

\begin{itemize}
\item The relevant fields
of the supersymmetry breaking MSB sector, of the MSSM sector, and of the messenger sector are
in thermal equilibrium. This is the case if the SUSY breaking scenario is gauge mediation,
direct mediation, or even a visible sector breaking. (On the other hand, a SUSY breaking sector which couples to MSSM
only gravitationally would remain out of thermal equilibrium.)

\item The MSB sector is IR free and the supersymmetry preserving vacua are induced by ``dynamical supersymmetry 
restoration''. This condition is enough to ensure that the SUSY preserving vacua $\vzero$ contain 
fewer light degrees of freedom than the metastable ground state near the origin, $\vplus$, because 
one always has to {\em integrate out} flavours to induce the dynamical term in the superpotential that restores 
the supersymmetric vacua. In addition, the dynamical restoration of supersymmetry guarantees a 
wide separation between $\vplus$ and $\vzero$ and hence the long lifetime of the metastable minimum.
\end{itemize}

In the present paper we will further refine and build upon this idea. There are two main results of this study.
First, we reconsider the lower bound on the reheat temperature in order for the 
Universe to be always driven to the non-supersymmetric MSB vacuum, $\vplus$, by thermal effects.
In ref.\cite{ACJK} a {\em sufficient} condition was obtained, namely that the reheat temperature should be 
higher than the vev of the supersymmetry preserving vacuum, $T_R\gtrsim \Phi_0 /\log \Phi_0^4$. This bound was derived 
by considering the effect of the $\Nf$ flavours of $\p$ and $\pt$ fields becoming heavy in the supersymmetric 
vacuum $\vzero$, but neglecting the effect of the mass-gap for gauge degrees of freedom 
generated by the confinement of $SU(N)$ in $\vzero$. Since the latter
would further reduce the number of light degrees of freedom in $\vzero$, it was argued 
that this could only lower the required reheat temperature. In this paper we determine by how much and find 
a surprising result; we show that although the mass-gap is parametrically small relative to $\Phi_0$, it in fact {\em dramatically} 
reduces the minimal value of the reheat temperature, $T_R$, necessary for the
Universe to end up in the non-supersymmetric $\vplus$. We will show that
$ T_R  >  {\rm few}\times \mu,$ where $\mu$ is the SUSY breaking scale, is sufficient. This result is also essentially independent
of the other parameters of the model, such as the position of the supersymmetry preserving vacua $\Phi_0 \gg \mu$ or the
numbers of colours and flavours $N$ and $\Nf$ of the ISS model.
To get an idea of the relevant scales
in our MSB-MSSM theory one can, for example, think of $\mu$ (the SUSY breaking scale) to be of the order
$\mu \sim\, 1 - 10^3 \, {\rm TeV}$ and $\Phi_0$ (the dynamically generated vev of the SUSY vacuum) to be
of the order e.g. $\Phi_0 \sim 100 \,\mu.$ Of course, a much wider range for both, $\mu$ and $\Phi_0$ is possible
and our analysis is  not affected by these values.

Our second observation, which we discuss in detail in Section 5, concerns the effect of the coupling of 
the MSB sector to the MSSM. In general one might be concerned that the MSSM sector could alter our picture drastically.
This is
because the mass-splittings between Standard Model particles and their superpartners
 induced in the metastable minimum can result in a large number of MSSM states 
being massive at the origin, but relatively light in the supersymmetry preserving minima.
This effect reduces the number of light degrees  of freedom in the vicinity of $\vplus$
relative to the SUSY preserving vacuum, and can potentially start eroding the minimum at the origin of
the thermal effective potential. 
However, we will show that due to the large separation between $\vplus$ and $\vzero$ in field space, and
under extremely general assumptions about the communication of the supersymmetry 
breaking to the visible sector and the induced mass-splittings, the entire visible sector 
is rendered harmless and the effects of its mass splittings essentially decouple from the discussion 
of the thermal and cosmological properties of the MSB sector.
Thus, once the Universe is driven to the SUSY breaking vacuum
$\vplus$ after reheating, it will not be able to escape back to the supersymmetric $\vzero$ even when we
couple the MSB sector to the MSSM sector and take into account the disappearance of mass splittings between 
partners and superpartners of the MSSM spectrum in $\vplus$. 

Two other interesting papers \cite{CFW,FKKMT} have appeared that are in accord with the conclusions 
of \cite{ACJK} and upon which 
the present discussion touches (although both \cite{CFW,FKKMT} began with the assumption that the 
system starts near the origin and were therefore somewhat orthogonal to our approach in \cite{ACJK}).  
Ref.\cite{CFW} discussed the fact that thermal effects 
stabilize the metastable minimum rather than drive it away from the origin, and our findings here 
confirm and extend that discussion: the coupling of the MSB sector to the MSSM cannot change that
conclusion for any reasonable choice of parameters. Ref.~\cite{FKKMT} showed in a careful analysis of the 
behaviour of the potential near the metastable origin, that on lowering the temperature to $\sim \mu$ 
there is a second order phase transition whose endpoint is the metastable minimum 
rather than the supersymmetric one. 

Other relevant recent work on metastable supersymmetry breaking in the context of strings, M-theory
and field theory includes \cite{F,O,B,F2,Forste,A2,Bena,Dine,Ahn,A,A3}.

\section{Recall: set-up for the ISS model and the effective potential}

We choose the MSB sector to be described by the simplest known theory which exhibits a metastable SUSY
breaking -- the original ISS model \cite{ISS}.  It is given by an
$SU(N)$ SYM theory coupled to $\Nf$ flavours of chiral superfields $\p$ and $\pt$ transforming in the
fundamental and the anti-fundamental representations of the gauge group.
There is also an
$\Nf \times \Nf$ chiral superfield $\Phi^i_j$ which is a gauge singlet.
The number of flavours is taken to be large, $\Nf  > 3N,$ such that the $\beta$-function for the gauge coupling is positive,
i.e.
$b_0 = 3N - \Nf < 0.$
The theory is free in the IR and strongly coupled in the UV where it develops a Landau pole at the energy-scale $\Lambda_L$.
At scales $E \ll \Lambda_L$ the theory is weakly coupled and
its low-energy dynamics as well as the vacuum structure is under control.
In particular, this guarantees a robust understanding of the theory
in the metastable SUSY breaking vacuum found in \cite{ISS}.\footnote{At energy scales of order 
$\Lambda_L$ and above, this effective description breaks down and one
should use instead a different (microscopic) description of the theory.
It is provided by its Seiberg dual formulation \cite{Seiberg1,Seiberg2} as explained in \cite{ISS}.
For our present purposes, we will always work at scales below  $\Lambda_L$ and we will only use the
macroscopic formulation of the ISS model.}

The tree-level superpotential of the ISS model is given by
\be
W_{\rm cl}\, =\, h\, \Tr_{\sst \Nf} \p \Phi \pt\, -\, h\mu^2\, \Tr_{\sst \Nf} \Phi
\label{Wcl}
\ee
where $h$ and $\mu$ are constants and $\mu$ is taken to be much smaller than the cut-off scale $\Lambda_L.$
The usual holomorphicity arguments imply that the superpotential
\eqref{Wcl} receives no corrections in perturbation theory. However, there is a non-perturbative contribution
to the full superpotential of the theory, $W=W_{\rm cl} + W_{\rm dyn},$ which 
is generated dynamically  \cite{ISS} and is given by
\be
W_{\rm dyn}\, =\, N\left( h^\Nf \frac{\det_{\sst \Nf} \Phi}{\Lambda_{L}^{\Nf-3N}}\right)^\frac{1}{N}
\label{Wdyn}
\ee
The authors of \cite{ISS} have studied the vacuum structure of the theory and established the
existence of the metastable vacuum $\vplus$ with non-vanishing vacuum energy $V_{+}$ characterised by
\be
\langle \p \rangle =\, \langle \pt^T \rangle = \, \mu \left(\begin{array}{c}
\uno_{N}\\ 0_{\Nf-N}\end{array}\right) \ , \quad
\langle \Phi \rangle = \, 0 \ , 
\qquad V_+ = \, (\Nf-N)|h^2 \mu^4| 
\label{vac+}
\ee 
as well as the SUSY preserving stable vacuum\footnote{In fact there are precisely $\Nf-N=\Nc$ of such vacua
differing by the phase $e^{2\pi i/(\Nf-N)}$ as required by the Witten index of the microscopic Seiberg dual formulation
of the ISS theory.}
$\vzero$,
\be
\langle \p \rangle =\, \langle \pt^T \rangle = \, 0 \ , \quad
\langle \Phi \rangle = \, \Phi_0=\, \mu \gamma_0 \, \uno_{\Nf} \ , \qquad \qquad V_0 = \, 0
\label{vac0}
\ee
where $V_0=0$ is the energy density in this vacuum and
\be
\gamma_0 =\,\biggl( h \epsilon^{\frac{\Nf-3N}{\Nf-N}}\biggr)^{-1} \ , \quad {\rm and} \quad
\epsilon :=\, \frac{\mu}{\Lambda_L} \ll \, 1.
\label{gamma0}
\ee
In the metastable nonsupersymmetric vacuum \eqref{vac+} the $SU(N)$ gauge group is Higgsed 
by the vevs of $\p$ and $\pt$ and the gauge degrees of freedom
are massive with $m_{\rm gauge} = g \mu.$ 
This supersymmetry breaking vacuum $\vplus$ originates from the so-called
rank condition, which implies that there are no solutions to the F-flatness
equation $F_{\Phi^j_i}=0$ for the classical superpotential $W_{\rm cl}.$ 
The appearance of the SUSY preserving vacuum
\eqref{vac0} is caused by the non-perturbative superpotential $W_{\rm dyn}$ and
can be interpreted in the ISS model as a non-perturbative
or dynamical restoration of supersymmetry \cite{ISS}.

In \cite{ACJK} we parameterised the path interpolating between the two vacua \eqref{vac+} 
and \eqref{vac0} in field space via 
\be
 \p(\sigma)  =\,  \pt^T (\sigma) = \,\sigma \mu\, \left(\begin{array}{c}
\uno_{N}\\ 0_{\Nf-N}\end{array}\right) \ , \quad
\Phi (\gamma) = \, \gamma \mu \uno_\Nf \ , 
\qquad  0 \le \gamma \le \gamma_0\ , \   1 \ge \sigma \ge 0
\label{inter}
\ee 
Since the Kahler potential in the free magnetic phase in the IR is that of the
classical theory, the zero-temperature effective potential $V_{T=0}(\gamma,\sigma)$ on the interpolating trajectory \eqref{inter}
can be determined directly from the superpotential of the theory 
\eqref{Wcl}, \eqref{Wdyn}.

Both, the interpolating trajectory, \eqref{inter}, and the resulting effective potential $V_{T=0}(\gamma,\sigma)$
are functions of two independent variables, $\sigma$ and $\gamma$, which parameterise 
the two directions in the field space: $\p=\pt= \sigma \mu$ and $\Phi=\gamma \mu.$ 
Instead of plotting
$V_{T=0}$ over the two-dimensional field-space, we have chosen in \cite{ACJK} to parameterise it by a single
combined direction as follows. For each fixed value of $\gamma$ we minimise $V_{T=0}$ is terms of $\sigma.$
For $0 \le \gamma \le 1$ the minimum is at $\sigma(\gamma)=\sqrt{1-\gamma^2},$
and for $1 \le \gamma$ the minimum is at $\sigma(\gamma)=0.$ The resulting expression
for the effective potential $V_{T=0}(\gamma,\sigma(\gamma))$ is now a function of $\gamma$ alone 
\begin{eqnarray}
 V_{T=0} (\gamma,\sigma(\gamma))
& = & {|h^2 \mu^4| }\, \left\{ \begin{array}{cc}
\Nf-N + \,2 N \gamma^2(1- \hf\gamma^2)  & \qquad 0 \le \gamma\le 1 \\
\\
 \Nf\biggl( \bigl(\frac{\gamma}{\gamma_0}\bigr)^{\frac{\Nf-N}{N}}-1\biggr)^2
& \qquad 1\leq\gamma\\
\end{array}\right.
\label{VzerodipT}
\end{eqnarray}
which we plot
in Figure 1 as the dashed line. 
\begin{figure}[t]
    \begin{center}
        \includegraphics[width=8cm]{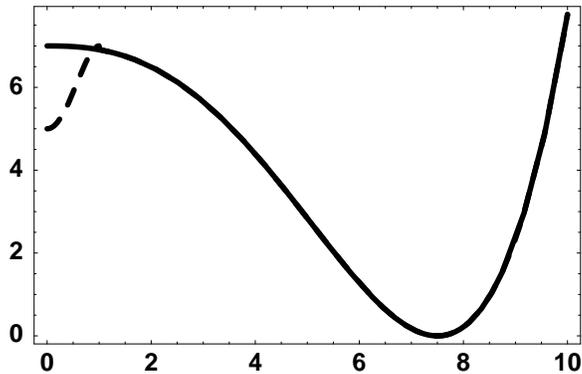}
    \end{center}
    \caption{Zero temperature effective potential $\hat{ V}_{T=0} (\gamma)$ as a function of 
    $\gamma = \Phi/\mu.$ Solid line denotes the potential of Eq.~\eqref{VzeroT}, and dashed line --
    the potential of Eq.~\eqref{VzerodipT}
    For the SUSY preserving vacuum $\vzero$ we chose $\gamma=\gamma_0=7.5$.
    The SUSY breaking metastable minimum $\vplus$ is always at $\gamma=0.$ 
    We have taken the minimal allowed values for $N$ and $\Nf$, $N=2$, $\Nf=7.$ }
    \label{fig:recurs}
\end{figure}
The SUSY preserving minimum $\vzero$ appears at $\gamma=\gamma_0$ far away from the origin as expected,
while the SUSY breaking minimum $\vplus$ shows up at the origin in Figure 1. 
However, from the two-dimensional $(\gamma,\sigma)$-perspective
$\vplus$ actually appears slightly away from the origin in the $\sigma = \p /\mu$ direction at
$\sigma = 1 \ll \gamma_0.$ When thermal effects are added to the effective potential, the values of
$V$ away from the origin will be lifted relative to the origin in the field space, and 
at temperatures above the restoration of $SU(N)$ temperature, the minimum
$\vplus$ will be shifted precisely to the origin  $\sigma =0$ and $\gamma=0.$
For this reason, in the present paper we will use for the zero-temperature effective potential
the expression without the dip at the origin, 
\be
V_{T=0} (\gamma)  =\, {|h^2 \mu^4| }\, \Nf\biggl( \bigl(\frac{\gamma}{\gamma_0}\bigr)^{\frac{\Nf-N}{N}}-1\biggr)^2 \ ,
\qquad 0\leq\gamma
\label{VzeroT}
\ee
This expression is plotted in Figure 1 as the solid line. We stress again, that at exactly zero temperature
the potential at the origin is at the saddle point and the metastable minimum $\vplus$ is reached in from there 
by stepping to $\sigma=1$ in the (orthogonal to the plot) $\sigma$ direction. However, at already sufficiently low
temperatures $T \ge T_{\sst SU(N)}$ the minimum shifts to the origin.

In Ref.~\cite{FKKMT} the $SU(N)$ restoration temperature was determined to be
\be
T_{\sst SU(N)} =\, \frac{\sqrt{2}\, \mu}{\sqrt{3\Nf-N} + {\cal O} (g^2)}  
\label{Trestor}
\ee
which for the minimal case of $\Nf=7$ and $N=2$ gives $T_{\sst SU(N)} \simeq 0.3\, \mu,$ and for all our applications
can be considered to be small.
The authors of \cite{FKKMT} have also argued that the phase transition from the minimum at the origin 
(at $T > T_{SU(N})$) to
the minimum $\vplus$ at $\sigma=1,$ $\gamma=0$ is second order. 
In the rest of the paper we will use the expression for the effective potential \eqref{VzeroT} which has no dip at the origin.
For $\Phi_0 \gg \mu$ or equivalently, $\gamma_0 \gg \sigma=1$ there will be little difference between the potentials in
\eqref{VzeroT}, in \eqref{VzerodipT} or the two-dimensional one.

The key features of this effective potential are (1) the large distance between the two vacua,
$\gamma_0 \gg 1,$ 
and (2) the slow rise of the potential to the left of the SUSY preserving vacuum.
(For aesthetic reasons $\gamma_0$ in Figure 1. is actually chosen to be rather small, $\gamma=\gamma_0=7.5.$)

The authors of \cite{ISS} have already estimated the tunnelling rate from the metastable $\vplus$
to the supersymmetric vacuum $\vzero$ by approximating the potential in Figure 1 in terms of a triangle.
It is always possible, by choosing sufficiently 
small $\epsilon$, to ensure that the decay time of the metastable vacuum 
to the SUSY ground state is much longer than the age of the Universe. 
On closer inspection the constraints imposed by this condition are 
in any case very weak.

\section{Effects of $\p$, $\pt$ and confining gauge fields at finite temperature}

The effective potential at finite
temperature along the $\Phi$ direction is governed by the following well-known expression \cite{Vthermal}:
\be
V_{T}(\Phi)=\, V_{T=0}(\Phi)+\frac{T^{4}}{2\pi^{2}}\sum_{i}\pm n_{i}\int_{0}^{\infty}\mbox{d}q\, 
q^{2}\ln\left(1\mp\exp(-\sqrt{q^{2}+m_{i}^{2}(\Phi)/T^{2}})\right)
\label{Vth1}
\ee
The first term on the right hand side is the zero temperature value of the effective potential.
The second term is the purely thermal correction (which vanishes at $T=0$) and it is determined 
at one-loop in perturbation theory. The $n_i$ denote the numbers of degrees of freedom present in the 
theory\footnote{Weyl fermions and complex scalars each count as $n=2.$} 
and the summation is over all of these degrees of freedom. The
upper sign is for bosons and the lower one for fermions. Finally, $m_{i}(\Phi)$ denote the masses of these degrees
of freedom induced by the vevs of the field $\Phi$.

As in \cite{ACJK} we will be using 
dimensionless variables for the field $\gamma = \Phi/\mu$ and temperature,
\be
\Theta =\, T/|h\mu|
\ee 
and will define a rescaled potential $\hat{V}$ which does not have the
overall constant $|h^2 \mu^4|$ (cf. Eq.~\eqref{VzeroT})
\be
\hat{V}_{\Theta=0} (\gamma)  =\, \frac{1}{|h^2 \mu^4| }\, {V}_{\Theta=0} (\gamma) \ , \quad
\hat{V}_{\Theta} (\gamma)  =\, \frac{1}{|h^2 \mu^4| }\, {V}_{\Theta} (\gamma)
\label{hatVTdef}
\ee

As we interpolate from the metastable vacuum to the supersymmetric one,
the $\Nf$ flavours of $\p$ and $\pt$ acquire masses $m_\p= h\Phi =h\mu \gamma$
and become heavy at large values of $\Phi.$ This effect gives the contribution to the second
term on the right hand side of \eqref{Vth1} of the form\footnote{
There are only two terms in the sum in \eqref{Vthphi}: one for bosons (+), and one for fermions (-). 
The prefactor $4N\Nf$ counts the total number of bosonic degrees of freedom in $\p^a_i$ and $\pt^i_a.$}
\be
\Delta \hat{V}_{\Theta\, \p,\pt}(\gamma)=\,
\frac{h^{2}}{2\pi^{2}}\Theta^{4}\sum_{\pm}\pm 4N \Nf \int_{0}^{\infty}\mbox{d}q\, 
q^{2}\ln\left(1\mp\exp(-\sqrt{q^{2}+\gamma^{2}/\Theta^{2}})\right).
\label{Vthphi}
\ee
The above contribution \eqref{Vthphi} to the effective potential takes into account the mismatch between the light
and heavy $\p$, $\pt$ degrees of freedom in the vicinity of the vacua $\vplus$ and $\vzero.$ This is
precisely the contribution already
taken into account in \cite{ACJK}.
We will now also include the effect of massive $SU(N)$ degrees of freedom (gauge fields and gauginos of the ISS sector).
In the SUSY breaking vacuum $\vplus$ these degrees of freedom can be counted as massless\footnote{
At very low temperatures the gauge group $SU(N)$ is Higgsed and gauge masses are $m_{\rm gauge}=g \mu$
which are considered negligible. At slightly higher temperatures  $T \sim \hf \mu$ the $SU(N)$ is restored
and all gauge degrees of freedom become truly massless.},
but in the
SUSY preserving vacuum $\vzero$ the gauge group is confined. This implies that there is a mass-gap
for the gauge degrees of freedom which is set by the value of the gaugino condensate $\langle \lambda \lambda \rangle.$

What happens is that away from the SUSY breaking $SU(N)$ vacuum $\vplus$ the mass-eigenstate spectrum for
gauge degrees of freedom changes from massless gauge fields and gauginos to massive colourless states.
We will model this effect by continuing to use the 1-loop approximation of \eqref{Vth1}, 
in terms of dressed or {\it massive} gauge fields and gauginos. Their mass is given by the mass-gap,
\be
m_{\rm gauge} = \, \langle \lambda \lambda \rangle^{\frac{1}{3}} = \,
\left(\frac{(h\mu\gamma)^{\Nf}}{{\Lambda_L}^{\Nf-3N}}\right)^{\frac{1}{3N}}
\label{mgap}
\ee
On the right hand side of \eqref{mgap} we used the (exact) value of the gaugino condensate as
determined by the nonperturbative superpotential \eqref{Wdyn} and we have expressed everything in terms
of the dimensionless variable $\gamma$. Relating to the value of $\gamma$ at the SUSY preserving vacuum $\vzero$
we rewrite \eqref{mgap} as 
\be 
m_{\rm gauge} = \, h^{\frac{1}{3}} \mu \left(\frac{\gamma}{\gamma_0}\right)^{\frac{\Nf}{3N}}
\gamma_0^{\frac{1}{3}}
\label{mgauge2}
\ee
which at large $\gamma$ lies in the expected range $ \mu \ll m_{\rm gauge} \ll m_\p$
for $\epsilon \ll 1$.

This analysis leads us to the following expression for the contribution of gauge degrees of freedom
to the thermal effective potential \eqref{Vth1}
\be
\Delta \hat{V}_{\Theta\, {\rm gauge}}(\gamma)=\,
\frac{h^{2}}{2\pi^{2}}\Theta^{4}\sum_{\pm}\pm 2(N^2-1) \int_{0}^{\infty}\mbox{d}q\, 
q^{2}\ln\left(1\mp\exp(-\sqrt{q^{2}+m_{\rm gauge}^{2}/T^{2}})\right)
\label{Vth2}
\ee
where 
\be
\frac{m_{\rm gauge}^{2}}{T^{2}} =\,  h^{-\frac{4}{3}} 
\left(\frac{\gamma}{\gamma_0}\right)^{\frac{2\Nf}{3N}}
\gamma_0^{\frac{2}{3}} \, \frac{1}{\Theta^2}
\label{mgTrat} 
\ee

\begin{figure}[t]
\begin{center}
\subfigure{
\begin{picture}(180,130)(-15,0)
\includegraphics[width=6.5cm]{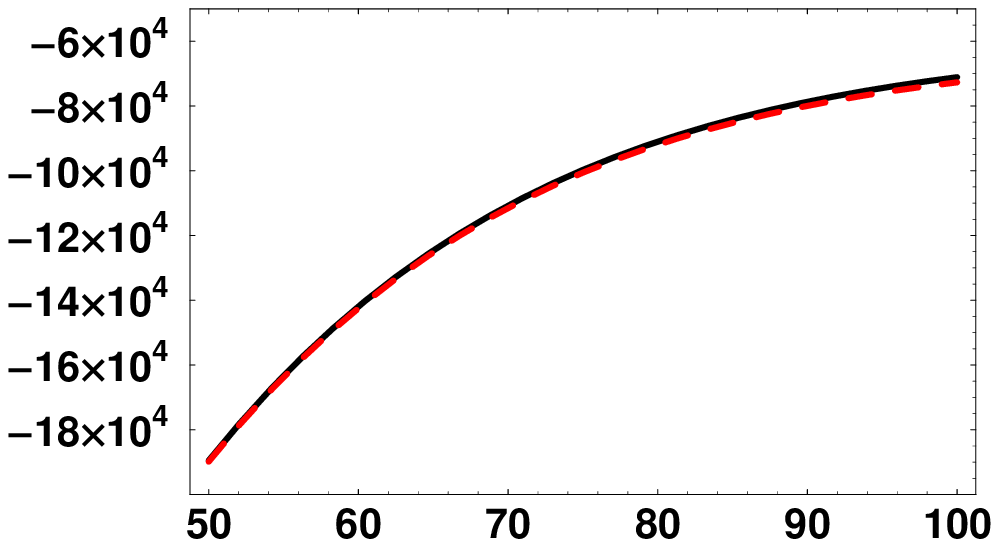}
\Text(-170,100)[l]{\scalebox{1.2}[1.2]{$\Theta=15\gg\Theta^{\rm{old}}_{\rm{crit}}$}}
\end{picture}
}
\hspace*{2.5cm}
\subfigure{
\begin{picture}(180,130)(0,0)
         \includegraphics[width=6.5cm]{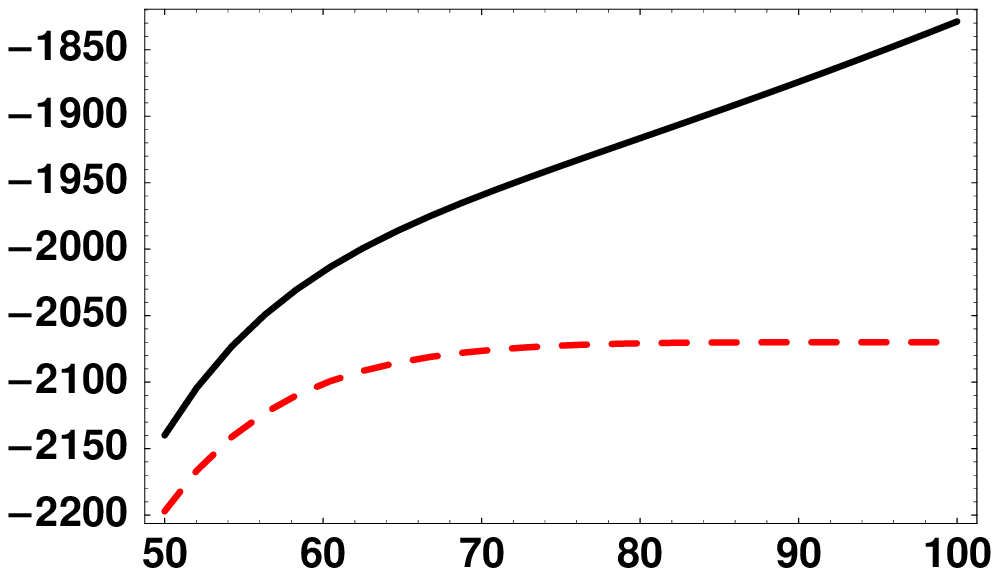}
\Text(-170,100)[l]{\scalebox{1.2}[1.2]{$\Theta=6.4\approx\Theta^{\rm{old}}_{\rm{crit}}$}}
\end{picture}
}
\subfigure{
\begin{picture}(180,130)(-15,0)
        \includegraphics[width=6.5cm]{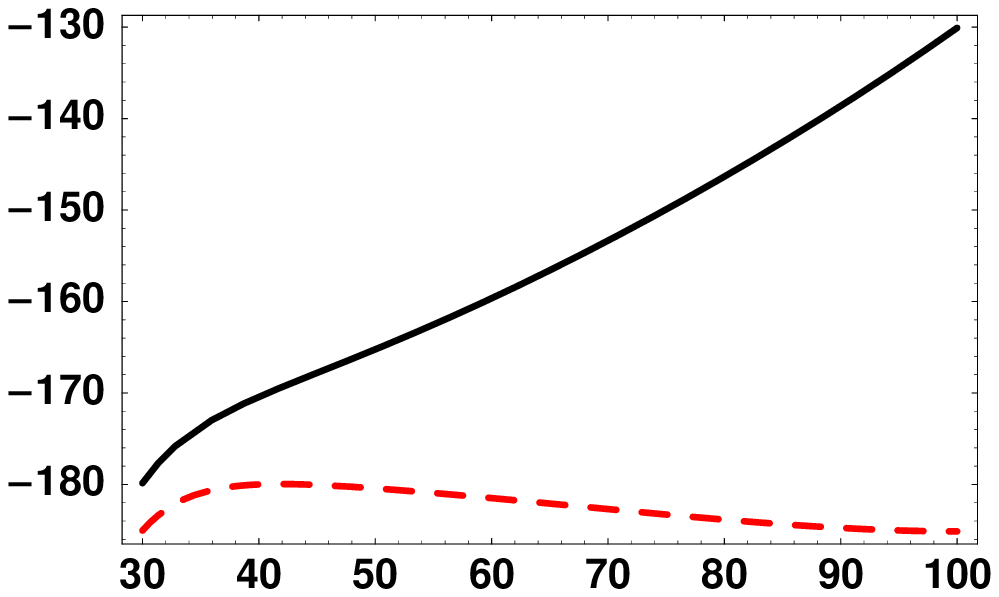}
\Text(-170,100)[l]{\scalebox{1.2}[1.2]{$\Theta=3.5>\Theta_{\rm{crit}}$}}
\end{picture}
}
\hspace*{2.5cm}
\subfigure{
\begin{picture}(180,130)(0,0)
        \includegraphics[width=6.5cm]{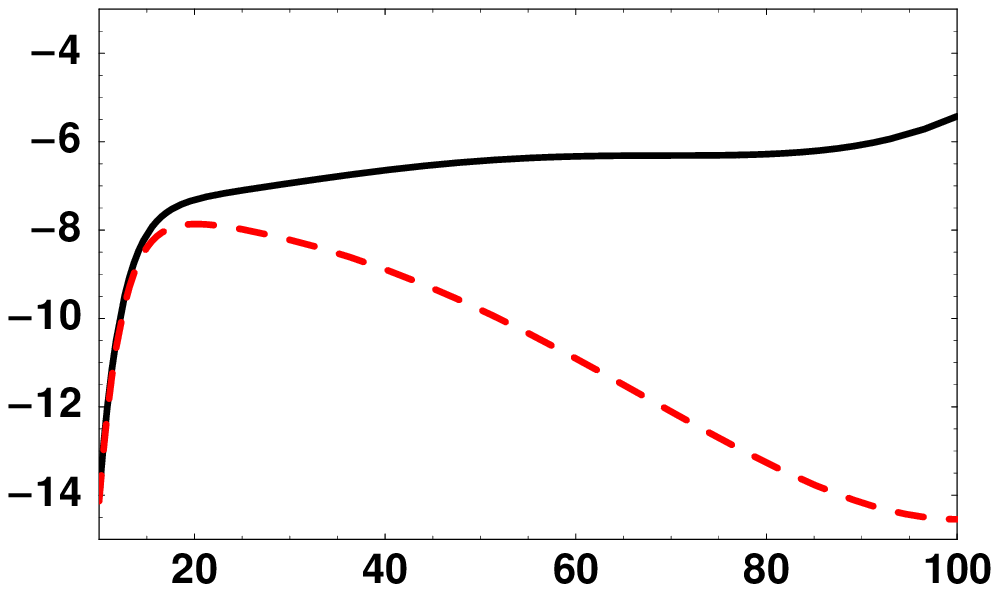}
\Text(-170,100)[l]{\scalebox{1.2}[1.2]{$\Theta=1.85\approx\Theta_{\rm{crit}}$}}
\end{picture}
}
\subfigure{
\begin{picture}(180,130)(-15,0)
        \includegraphics[width=6.5cm]{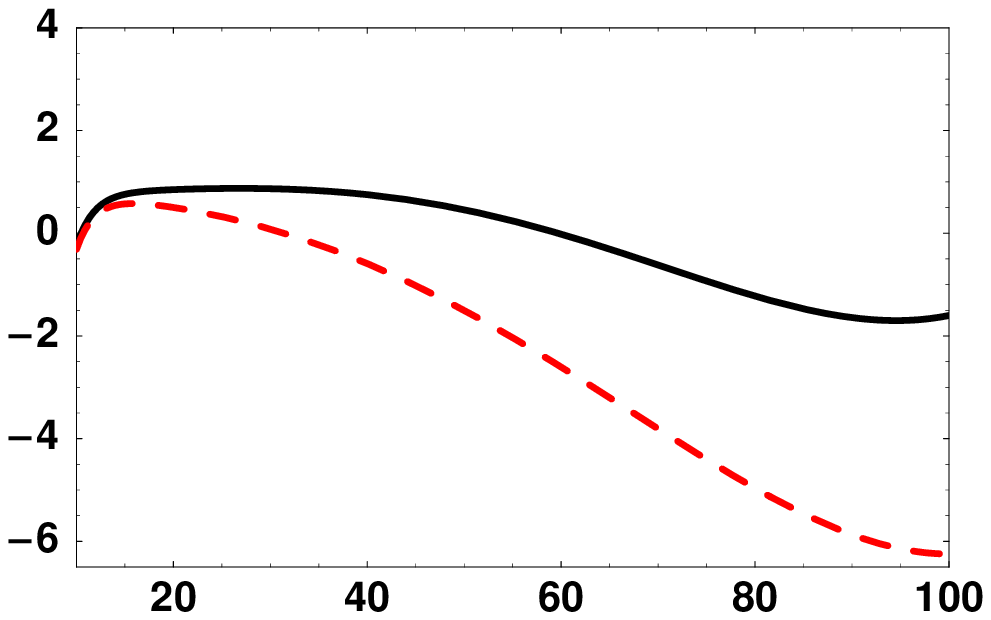}
\Text(-170,100)[l]{\scalebox{1.2}[1.2]{$\Theta=1.5<\Theta_{\rm{crit}}$}}
\end{picture}
}
\hspace*{2.5cm}
\subfigure{
\begin{picture}(180,130)(0,0)
\includegraphics[width=6.5cm]{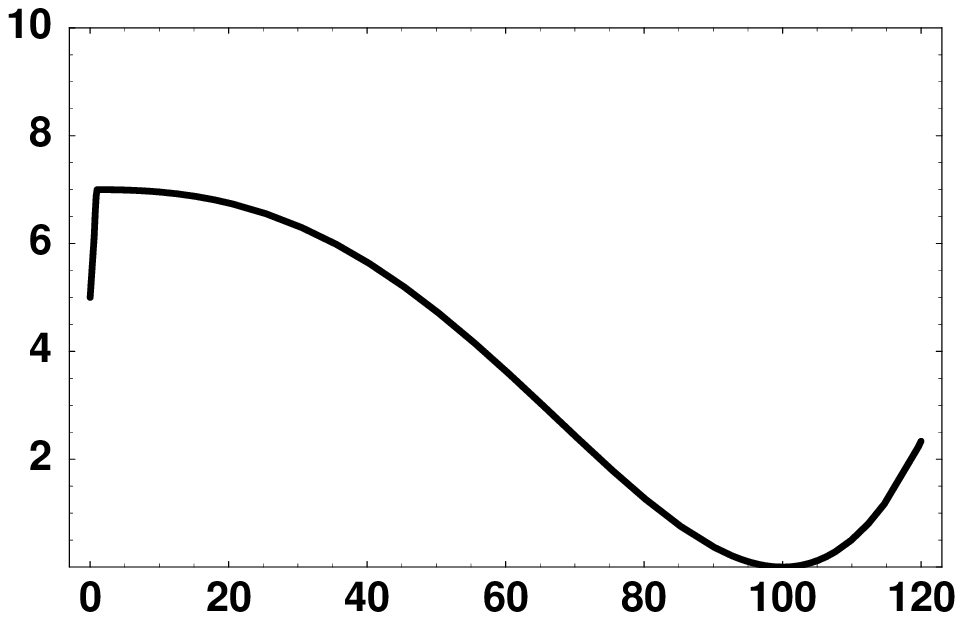}
\Text(-170,100)[l]{\scalebox{1.2}[1.2]{$\Theta=0$}}
\end{picture}
}
\end{center}
    \caption{Potential for different values of the temperature $\Theta$.
    The solid black line includes thermal corrections from the $\varphi$-fields as well as the gauge bosons, 
    whereas for the red dashed line we have neglected the masses of the gauge bosons. We choose $\gamma_0=100$ 
    and the minimal case $\Nf=7$ and $N=2.$ 
     }
    \label{combined}
\end{figure}

In Fig. \ref{combined} we plot the thermal effective potential of the ISS model including gauge mass-gap
effects (black solid line)
\be
\hat{V}_{\Theta \, \rm tot} =\, \hat{V}_{\Theta=0}  (\gamma) +
\Delta \hat{V}_{\Theta\, \p,\pt}(\gamma) + \Delta \hat{V}_{\Theta\, {\rm gauge}}(\gamma)
\label{Vtot}
\ee
versus the expression for $\hat{V}_{\Theta}$ where the mass-gap for gauge degrees of freedom is not
taken into account (red dashed line),
\be
\hat{V}_{\Theta} =\, \hat{V}_{\Theta=0}  (\gamma) +
\Delta \hat{V}_{\Theta\, \p,\pt}(\gamma) + \Delta \hat{V}_{\Theta\, {\rm gauge}}(\gamma=0)
\label{Vtotapp}
\ee
The last term in the above equation, $\Delta \hat{V}_{\Theta\, {\rm gauge}}(\gamma=0),$ is a constant shift of the potential as a whole;
it does not affect the relative heights of $\vplus$ and $\vzero$ and is included to represent the contributions of massless
gauge degrees of freedom.

\begin{figure}[t]
\begin{center}
\subfigure{       
\begin{picture}(180,130)(0,0)
\includegraphics[width=6.5cm]{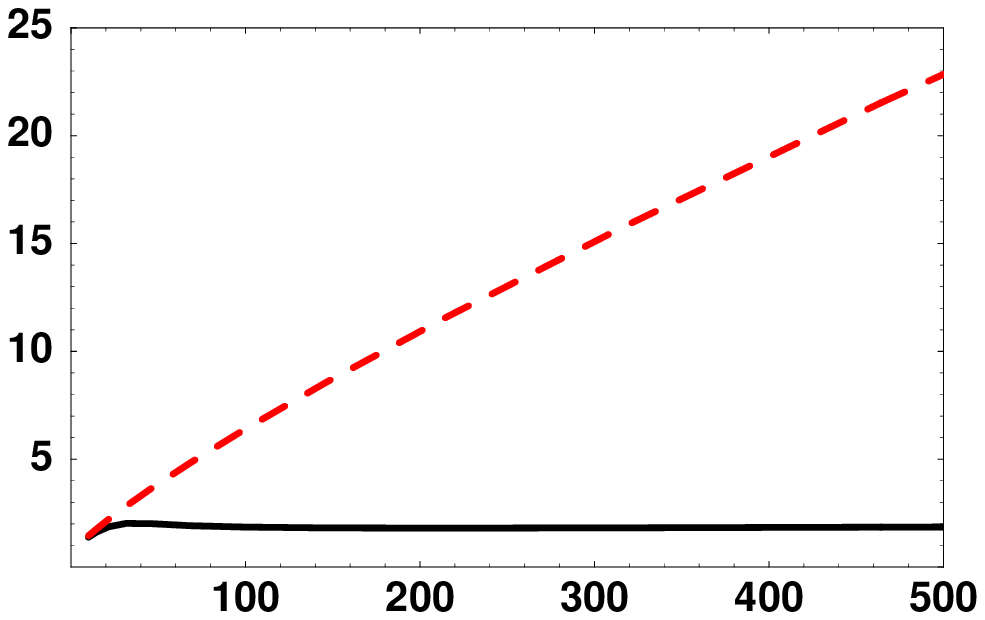}
\end{picture}
}
\hspace{2cm}
\subfigure{
\begin{picture}(180,130)(0,0)        
\includegraphics[width=6.5cm]{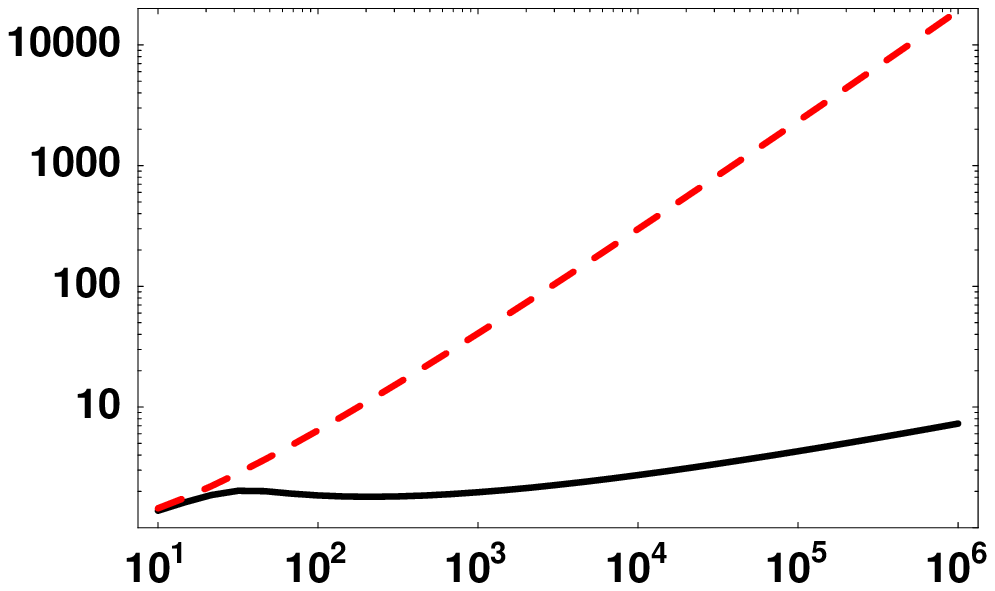}
\end{picture}
}
\end{center}
    \caption{$\Theta_{\rm crit}$ (solid black) and $\Theta_{\rm crit}^{\rm old}$ (red dashed) 
    as functions of $\gamma_0$ for $\Nf=7$ and $N=2$. The left panel is plotted linearly  whereas the right panel is
    a double logarithmic plot. 
     }
\label{compare}
\end{figure}

It follows from our results in Fig.~\ref{combined} that the inclusion of the dynamical gauge boson masses 
in \eqref{Vtot} does lower 
the critical temperature significantly. This is even more apparent in Fig.~\ref{compare} where we compare the critical 
temperature with gauge boson masses included to the one obtained when we include only the 
mass effects of $\varphi$ and $\tilde{\varphi}$ fields.
Indeed, performing a similar analysis as the one in section 3.2.1 of Ref.~\cite{ACJK} one obtains in leading order for large $\gamma_{0}$,
\begin{equation}
 \Theta_{\rm{crit}}\sim \frac{\gamma^{\frac{1}{3}}_{0}}{\log\left(\gamma^{\frac{4}{3}}\right)},
\end{equation}
which is parametrically smaller than $\gamma_{0}$. For not too large values of $\gamma_0$, say $\gamma_{0}\lesssim 1000,$
we note (cf. Fig. \ref{compare}) that the resulting critical temperature is of order a few $\mu$ and does not depend strongly
on the value of $\gamma_0.$ 

\begin{figure}[t]
\begin{center}
\subfigure{  
\begin{picture}(180,130)(0,0)
\includegraphics[width=6.5cm]{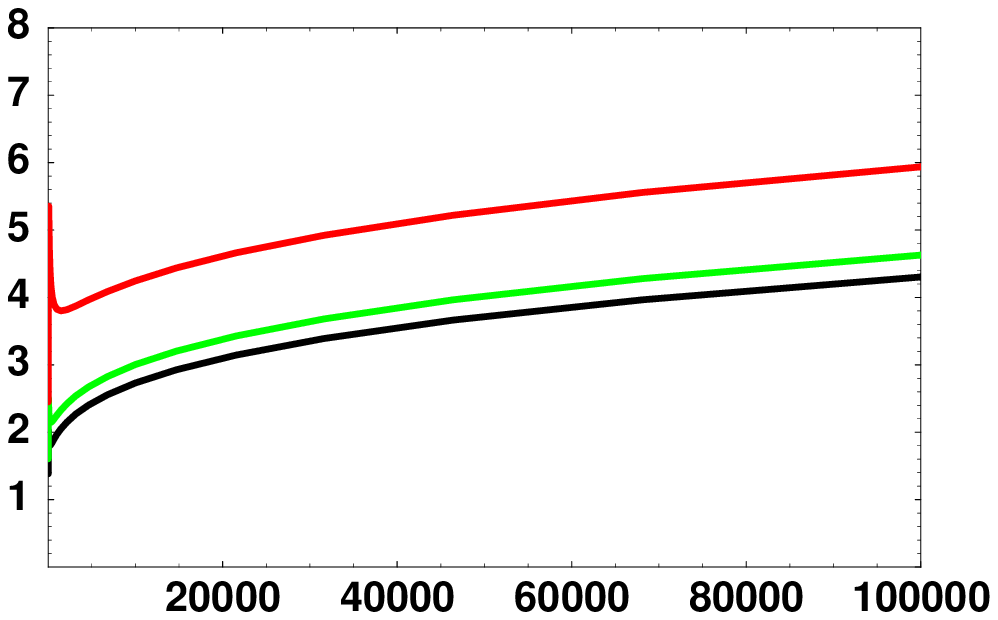}
\end{picture}
}
\hspace{2cm}
\subfigure{
\begin{picture}(180,130)(0,0)
\includegraphics[width=6.5cm]{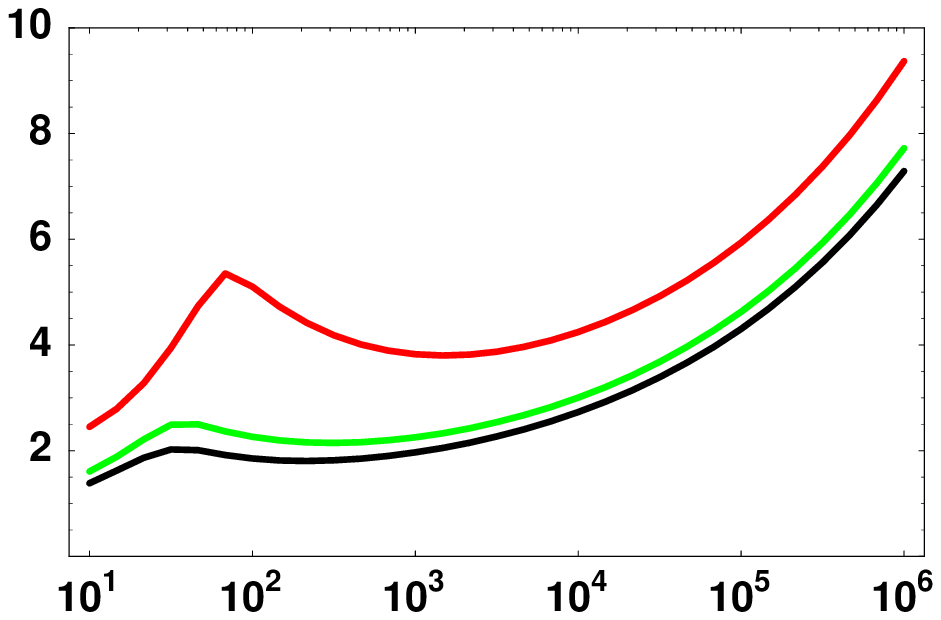}
\end{picture}
}
\end{center}
\caption{$\Theta_{\rm crit}$ as a function of $\gamma_0$ for
    $\Nf=7$ (black),  $\Nf=12$ (green), and $\Nf=100$ (red). The number of colours is $N=2$.
    The left plot is linear whereas the right is logarithmic on the horizontal axis. 
    Note that the dependence of $\Theta_{\rm{crit}}$ on $\gamma_{0}$ is always relatively weak.
     }
 \label{various}
\end{figure}

Furthermore $\Theta_{\rm crit}$ is not very sensitive to the choice of $\Nf$ and $N$ either,
as we can see in Fig. \ref{various}.

\section{Behaviour of the field after nucleation/rolling}

As in \cite{ACJK}, one should examine the behaviour of the fields to make
sure that for temperatures $T \gtrsim T_{\rm crit}$
the phase transition to the vacuum at the origin completes. 
There are two timescales that
are of concern: the time for the field $\Phi$ to roll to $|\mbox{vac}_{+}\rangle$
and the time for oscillations around $|\mbox{vac}_{+}\rangle$ to
be damped. The situation can in principle be much more delicate than anticipated
in our previous paper since as we have seen in the previous section, the mass scales involved
are parametrically different. In this section we will therefore
briefly revisit the cosmological discussion in order to make sure
our previous conclusion that at temperatures $T \gtrsim T_{\rm crit}$ the phase transition completes is unchanged.

Again, in order to make order of magnitude estimates of timescales,
it is sufficient to model the classical potential as linear,
\begin{equation}
V  =  \, {\rm const}\, T^{4} \frac{\Phi}{\Phi_{0}}.
\end{equation}
Neglecting for the moment the damping effect of the $\varphi\Phi\tilde{\varphi}$
coupling, and solving the field equations assuming that the contribution of the Hubble constant, $H\sim T^{2}/M_{\rm Pl}$
is negligible, we find that the field falls to the origin in time
\begin{equation}
\Delta t=\, {\rm const}^{\prime}\,\, \frac{\Phi_{0}}{T_{0}^{2}}\sim\, \frac{\Phi_{0}}{M_{\rm Pl}}\, t_{0},
\end{equation}
where we have assumed a radiation dominated Friedmann-Robertson-Walker universe with scale factor
$a(t)\sim (t/t_{0})^{\frac{1}{2}}$ and $T=T_{0}/a(t)$.

This time $\Delta t$ should be much shorter than the time needed for the Universe to cool enough
for the origin to be lifted, which is $t_{\rm cool}\sim t_{0}.$ This condition is satisfied 
\be
t_{\rm cool} \sim \, t_0 \gg \, \Delta t \ , \qquad {\rm as \, long\, as} \, \Phi_{0}\ll M_{Pl}
\ee
which in our settings is always the case.

After the transition, the field $\Phi$ undergoes coherent oscillation
about the origin. Again, the damping provided by the $h\varphi\Phi\tilde{\varphi}$
coupling captures the field $\Phi$ at the origin and prevents it escaping
as $T$ cools: the decaying oscillation amplitudes are of order
\be
\frac{\Phi_{\rm max}(T)}{\Phi_{0}} =\, \sqrt{\frac{T}{T_{0}}}\, e^{-\frac{1}{2}\Gamma_{\Phi}(t-t_{0})}
\ee
where the typical decay rate is $\Gamma_{\Phi}\sim T$. As a most
conservative case, assuming that the initial temperature was of order
$\mu$ and initial oscillations of order $\Phi_{0}$, we use
\be
T \sim\,  T_0 \sim\,  \mu \ , \qquad 
\Gamma_{\Phi}(t-t_{0}) \sim \, \mu \, t_0 \sim \, \mu\, M_{\rm Pl}/T^2 \sim \, M_{\rm Pl}/\mu
\ee
and find
\be
\frac{\Phi_{max}}{\mu}\sim \,
\frac{\Phi_{0}}{M_{Pl}}\frac{M_{Pl}}{\mu}e^{-\frac{1}{2}\frac{M_{Pl}}{\mu}} \le \, 2e^{-1}\frac{\Phi_{0}}{M_{Pl}}
< \, 1
\ee
or $\Phi_{max} < \mu \ll \Phi_0$ and the oscillations are damped.

\section{Effects of Standard Model mass splittings}

So far we have considered thermal effects in the pure ISS model. In a more realistic scenario the ISS model will  
act as the supersymmetry breaking sector coupled to a supersymmetric extension of the Standard Model.
In this section we will include effects of the Standard Model particles and their superpartners.

In the metastable state $\vplus$ supersymmetry is broken at a scale $\sim \mu$. In the supersymmetric state $\vzero$ 
supersymmetry is unbroken. This implies that the mass splittings between SM particles and their superpartners 
are zero in $\vzero$, and 
of the order $\alpha \mu$ in $\vplus$ (where $\alpha$ is a coupling for the mediation of SUSY breaking;
in the following analysis we assume $\alpha^{2}<1$). 

\begin{figure}[t]
    \begin{center}
\includegraphics[width=8cm]{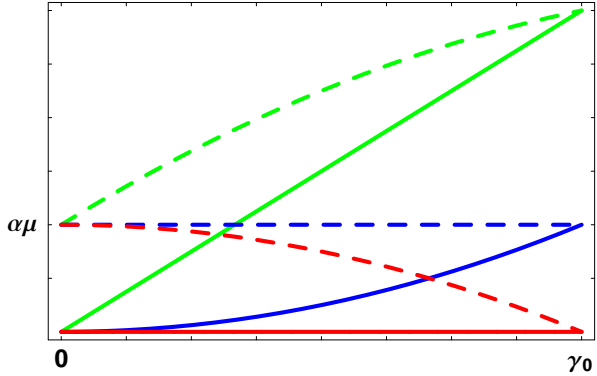}
    \end{center}
\caption{Possible scenarios for the masses of the SM particles (solid) and their superpartners (dashed) as a function of $\gamma$. 
In the state $\vplus$ at $\gamma=0$ SM particles are light and superpartners have masses $\alpha\mu$. 
At $\gamma=\gamma_{0}$ supersymmetry is restored and the difference between the masses of particles and their 
superpartner vanishes. Away from the special points $\gamma=0$ and $\gamma=\gamma_{0}$ the behavior is model dependent. 
The different colours depict three possibilities how the masses could behave.}
    \label{schematic}
\end{figure}

In principle we can envision several possible ways that SUSY can be restored in $\vzero$
(in the following we neglect all SM masses in $\vplus$ whereas all superpartners get masses $\alpha\mu$ in $\vplus$). 
In Fig. \ref{schematic} we schematically depict three possibilities for how the masses of SM particles and their 
superpartners behave as functions of $\gamma$. If both, the masses of the SM particles and their superpartners, 
are monotonically increasing (green and blue curves in Fig. \ref{schematic}) they act in the same direction as the 
$\varphi$-fields whose masses also increase with $\gamma$. At nonvanishing temperature they will stabilize $\vplus$. 
However, if the masses decrease towards $\gamma_{0}$ (red curves in Fig. \ref{schematic}) we expect a destabilising 
effect, because we get additional massless degrees of freedom in the supersymmetric vacuum $\vplus$. 
Let us now investigate whether this destabilizing effect can lead to transitions towards the supersymmetric vacuum.

We expect the strongest destabilizing effect on $\vplus$ if the masses of the superpartners tend to zero at $\gamma=\gamma_{0}$ 
while the masses of the SM particle remain zero (red line in Fig. \ref{schematic}). We consider two simple possibilities
for the MSSM sector superpartner masses $m_{SP}$,
\begin{eqnarray}
\label{masses1}
m_{SP}(\gamma)\!\!&=&\!\!\alpha\mu\frac{(\gamma_{0}-\gamma)}{\gamma_{0}}, 
\\
\label{masses2}
m^{\prime}_{SP}(\gamma)
\!\!&=&\!\!\alpha\mu\left(\frac{V_{T=0}(\gamma)}{V_{T=0}(0)}\right)^{\frac{1}{4}}
=\alpha\mu\left(1-\left(\frac{\gamma}{\gamma_{0}}\right)^{\frac{\Nf-N}{N}}\right)^{\frac{1}{2}}.
\end{eqnarray}
The first one, $m_{SP}$ is the simplest possible linear behavior whereas $m^{\prime}_{SP}$ relates the supersymmetry 
breaking to the value of the potential, Eq. \eqref{VzeroT}.

Transitions to the supersymmetric vacuum can only occur for temperatures below the degeneracy temperature $T_{\rm degen}$. 
In general $T_{\rm degen}$ will be shifted to slightly larger values in the presence of particles which are massless at 
$\gamma_{0}$ but massive at\footnote{For the same reason $T_{\rm crit}$ will slightly increase, too.} $\gamma=0$.  
\begin{equation}
 T^{SM}_{\rm degen}=T_{\rm degen}
 +\frac{1}{16}\left(\frac{3}{2\pi^{2}}\right)^{\frac{3}{4}}\Nf^{-\frac{1}{4}}(2N\Nf+N^{2}-1)^{-\frac{3}{4}}
 \frac{\alpha^{2}}{h^{\frac{1}{2}}}
 N_{SM}\mu
 +{\mathcal{O}}\left(\alpha^4N^{2}_{SM}\right),
\end{equation}
where $N_{SM}$ is the number of massive superpartners half of which are bosons and the other half are fermions.
In the following we will assume\footnote{For too small $T_{\sst SU(N)}$, the stabilization of the local minimum 
$\sim - T^{4}$ at the origin becomes very weak just above $T_{\sst SU(N)}$ and tunnelling in the direction of $\vzero$ may become 
possible for relatively small values of $\gamma_{0}$ even in absence of additional SM sector particles. 
The inequality $T_{\sst SU(N)}>0.2$ places a rather mild bound on $N_{f}$. For small gauge coupling $g^{2}\ll h^{2}$, $3N_{f}-N<50$ 
is a sufficient condition.} 
$T_{\rm degen}>T>T_{\sst SU(N)}>0.2\,\mu$. For temperatures below $T_{\sst SU(N)}$ the supersymmetry breaking minimum is even more 
stable due to the formation of the $\sigma$ condensate (i.e. Higgsing  of the $SU(N)$).

Let us first determine whether the SM particles can completely erode the minimum at the origin thereby allowing for classical 
transitions to the SUSY vacuum.
For $T>T_{\sst SU(N)}$ and close to $\gamma=0$ the zero-temperature contribution $V_{T=0}$, Eq. \eqref{VzeroT}, to the potential is
\begin{equation} 
\label{zerotilde}
V_{T=0}(\gamma)\approx \Nf\left(1-2\left(\frac{\gamma}{\gamma_{0}}\right)^{\frac{\Nf-N}{N}}\right) \, |h^2 \mu^4|.
\end{equation}
As long as the masses are $m\lesssim T$ the thermal contribution gives,
\begin{equation}
\label{close}
 \Delta V_{T}=-\frac{\pi^2}{90}T^{4}\left(n_{B}+\frac{7}{8}n_{F}\right)
 +\frac{1}{24}T^{2}\left(\sum_{B} m^{2}_{B}+\frac{1}{2}\sum_{F}m^{2}_{F}\right)
\end{equation}
 where $n_{B}$ and $n_{F}$ are the number of fermionic and bosonic degrees of freedom and $m_{B}$ and $m_{F}$
 correspond to the masses of bosons and fermions. The masses of the $\varphi$-fields and the gauge boson masses 
 grow with $\gamma$ and tend to stabilize a local minimum at $\gamma=0$ (cf. Eq. \eqref{close}). 
 The masses from the SM sector can, however, decrease (red line in Fig. \ref{schematic}) thereby destabilizing $\gamma=0$. 
 Neglecting the gauge boson masses for simplicity and assuming that all massive particles of the SM sector have 
 equal masses $m_{SM}$ the thermal contribution is,
 \begin{equation}
  \Delta \hat{V}_{\Theta}(\gamma)=-\frac{\pi^2}{96}h^{2} \Theta^{4}\left(8 N \Nf+N_{SM}\right)
  +\frac{N\Nf}{4}h^{2}\Theta^{2}\gamma^{2}+\frac{N_{SM}}{32}\Theta^{2}m^{2}_{SM}(\gamma), 
  \quad {\rm{for}}\quad m_{SM},\,h^{2}\gamma\lesssim T.
\end{equation}
If we can find a point $\gamma_{1}<\gamma_{0}$ such that
\begin{equation}
V_{\Theta}(\gamma_{1})>V_{\Theta}(0)
\end{equation}
we have to cross a barrier on our way to the supersymmetric minimum at $\gamma_{0}$ and classical evolution to the state $\vzero$ is impossible.
Using Eq. \eqref{masses1} for the masses of the massive SM sector fields we find
\begin{equation}
 \hat{V}_{\Theta}(1/10)-\hat{V}_{\Theta}(0)
=\frac{N\Nf}{400}\Theta^{2}h^{2}\mu^{2}\left(1-\frac{5}{2}\frac{N_{SM}}{h^{2}N\Nf}\frac{\alpha^{2}}{\gamma_{0}}\right)
+{\mathcal{O}}\left(\gamma^{\frac{N-\Nf}{N}}_{0}\right).
\end{equation}
If the term in the brackets is positive we have the desired barrier. For $h\sim 1$, $\gamma_{0}\sim 100$ and 
$N_{SM} < 10^{3}$ this allows all $\alpha^{2}\lesssim 0.5$. 
Using the for Eq. \eqref{masses2} for the mass we obtain
\begin{equation}
 \hat{V}_{\Theta}(1/10)-\hat{V}_{\Theta}(0)
=\frac{N\Nf}{400}\Theta^{2}h^{4}\mu^{2}+{\mathcal{O}}\left(\gamma^{\frac{N-\Nf}{N}}_{0}\right)>0.
\end{equation}
In this order we get no bound on $\alpha$ at all because the dependence of the mass on $\gamma_{0}$ 
is even weaker in the interesting region around the origin. 

\begin{figure}[t]
    \begin{center}
\includegraphics[width=8cm]{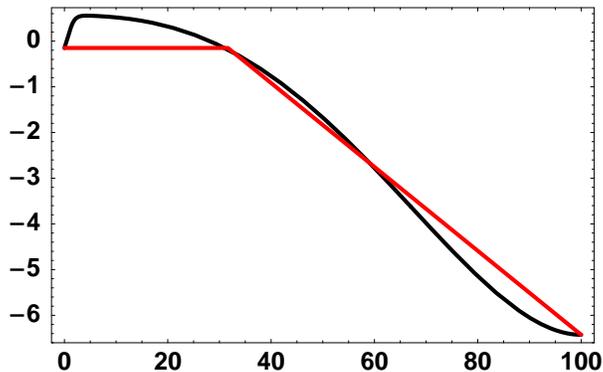}
    \end{center}
\caption{Flat potential (red) used in the model the true potential (black) in the estimate of the tunnelling rate.}
    \label{tunneling}
\end{figure}
 
Let us now check whether tunnelling is sufficiently suppressed. For the tunnelling we model the potential by a 
flat potential depicted in Fig. \ref{tunneling}. Between $\gamma=0$ and $\gamma=\gamma_{x}$, 
the potential is flat with value $V_{T}(\gamma_{x})=V_{T}(0)$. Then it decreases linearly towards $V_{T}(\gamma_{0})$ at $\gamma_{0}$.
Using similar approximations as above\footnote{For 
$h^{2}\Theta^{2}\lesssim\gamma^{2}_{x}/\gamma^{2}_{0}$ the change in the zero temperature contribution, 
Eq. \eqref{zerotilde}, to the potential cannot be neglected. We have checked numerically that the relation 
nevertheless holds in these cases, too.} one finds that 
\begin{equation}
\frac{\gamma_{x}}{\gamma_{0}}\gtrsim 0.05 \quad {\rm{for}} \quad \gamma_{0}\gtrsim 50, \,\,\,N_{SM}\lesssim 10^{3}
\quad {\rm{and}} 
\quad \alpha\lesssim 0.1,
\end{equation}
demonstrating that the thickness of the barrier grows parametrically with $\gamma_{0}$.
For our model potential the bounce action is (cf. \cite{ACJK}),
\begin{equation}
 \frac{S^{3D}}{T}=\frac{8\pi}{\Theta h^{2}}\frac{\sqrt{3}}{5}\Nf
 \gamma^{\frac{5}{2}}_{x}(\gamma_{0}-\gamma_{x})^{\frac{1}{2}}
 \end{equation}
and
\begin{equation}
\frac{S^{3D}}{T}\gtrsim 250  \quad {\rm{for}}\quad  \gamma_{0}\gtrsim 50.
\end{equation}

We conclude that under relatively mild conditions, SM sector particles do not induce transitions to the supersymmetric
vacuum $\vzero$. 

\section{Conclusions}

In this paper and in its companion \cite{ACJK} we have examined the dynamics of 
metastable SUSY breaking vacua in a cosmological setting. 
We conclude that generically the Universe is driven to the supersymmetry breaking
metastable vacuum by thermal effects.

In this paper we focused on the contributions to the effective potential of the gauge
degrees of freedom which develop a mass-gap in the SUSY preserving confining vacua in the MSB sector.
This reduces drastically the temperature required for the Universe to be driven to the 
metastable vacuum. We found that essentially any temperature larger 
than a few times 
the supersymmetry breaking scale $\mu$ is sufficient to ensure that the Universe ends in the
desired nonsupersymmetric vacuum state.

Furthermore, we have investigated the effects of the MSSM sector fields on the fate of this
nonsupersymmetric vacuum. We have found  (under very mild assumptions) that they are 
negligible and do not affect the stability of $\vplus.$

\section*{Acknowledgements}   

We thank Chong-Sun Chu, Stefan Forste and George Georgiou for useful discussions.
VVK is supported by a PPARC Senior Fellowship.

\end{document}